\documentclass{article}
\usepackage{multicol}
\usepackage{multirow}
\usepackage{booktabs}
\usepackage{amsmath}
\usepackage{amssymb}
\usepackage{amsfonts}
\usepackage{mathabx}
\usepackage{tablefootnote}
\usepackage{spconf}
\usepackage{graphicx}
\usepackage{hyperref}
\def\figureautorefname~#1\null{Fig.~#1\null}

\title{Generalized Zero-shot Audio-to-Intent Classification}

\name{Veera Raghavendra Elluru, Devang Kulshreshtha, Rohit Paturi, Sravan Bodapati, Srikanth Ronanki}
\address{
    AWS AI Labs
}
    
\copyrightnotice{979-8-3503-0689-7/23/\$31.00~\copyright2023 IEEE}
\begin{document}
%
\maketitle
\begin{abstract}

Spoken language understanding systems using audio-only data are gaining popularity, yet their ability to handle unseen intents remains limited. In this study, we propose a generalized zero-shot audio-to-intent classification framework with only a few sample text sentences per intent. To achieve this, we first train a supervised audio-to-intent classifier by making use of a self-supervised pre-trained model. We then leverage a neural audio synthesizer to create audio embeddings for sample text utterances and perform generalized zero-shot classification on unseen intents using cosine similarity. We also propose a multimodal training strategy that incorporates lexical information into the audio representation to improve zero-shot performance. Our multimodal training approach improves the accuracy of zero-shot intent classification on unseen intents of SLURP by 2.75\% and 18.2\% for the SLURP and internal goal-oriented dialog datasets, respectively, compared to audio-only training.

\end{abstract}

\vspace{-0.3cm}
\section{Introduction}
\label{sec:intro}
\vspace{-0.3cm}


The use of audio-based systems such as virtual assistants and voice-controlled devices \cite{lopez2018alexa, pearl2016designing}, has dramatically increased in popularity over the past few years. The effectiveness of these systems largely depends on their ability to accurately decipher a speaker's intent to provide suitable responses and execute requested actions. Audio-to-intent classification, an essential part of these systems, involves categorizing audio utterances into predetermined intent categories, allowing audio systems to perform specific tasks more efficiently.


Some approaches have been proposed in the past to perform audio-to-intent classification \cite{haghani2018audio, chen2018spoken, ray2021listen}. However, they require large amounts of data, typically in the order of several thousands of utterances, limiting their applicability in the real world. Pre-trained self-supervised learning (SSL) methods using Transformer architecture \cite{vaswani2017attention} such as HuBERT \cite{hsu2021hubert} and Wav2Vec \cite{schneider2019wav2vec} have shown promising results in various speech-related tasks such as Automatic Speech Recognition \cite{hsu2021hubert}, Emotion Recognition \cite{srinivasan2022representation}, and Speech Translation \cite{zhang2021transformer} with minimal amount of training data for each task. However, their exploration of audio-to-intent classification remains limited \cite{lai2021semi, yang2021superb}. 

Moreover, spoken language understanding systems that perform audio-to-intent classification \cite{hubert_wav2vec, higuchi2022bert} assume a fixed set of intent classes during both training and inference. This constraint restricts their applicability in real-world scenarios. Zero-shot Learning \cite{palatucci2009zero} can be leveraged for audio-to-intent classification to bridge this limitation. However, despite many works on zero-shot classification in computer vision \cite{frome2013devise, xian2018feature, wang2018zero, han2021contrastive}, a limited amount of works exist in the audio processing field. Prior works such as \cite{xie2021zero, primus2022improved} typically propose learning a compatibility function to map audio and class label text in the same embedding space during training, while projecting audio into text space allows classification of audio samples into unseen classes using their textual representations during inference. Another work by \cite{islam2019soundsemantics} follows a similar approach, but additionally proposes a two-way classifier based on a threshold-based decision algorithm to identify if the audio belongs to seen or unseen classes. We also note that the above methods are not specifically for \emph{intent} classification but general audio classification.

The problem of zero-shot classification can be further generalized to predict both seen and unseen intents from the training set. This is often the case in a conversational AI bot system where the full list of intents is not available during training. This scenario is referred to as Generalized Zero-shot Classification (GZSC) \cite{felix2018multi}. While considerable research has been conducted on GZSC models for textual and image classification \cite{siddique2021generalized, mahapatra2021medical}, there has been a lack of exploration in the realm of audio-based intent classification.

Previous approaches \cite{hubert_wav2vec, higuchi2022bert} make use of audio-only data for audio-to-intent classification. Various audio classification tasks, such as emotion recognition, have been shown to benefit from multimodal data. For example, a multimodal teacher-student fusion model is proposed in \cite{srinivasan2022representation} for improving audio-based emotion recognition. Similarly, \cite{huzaifah2023analysis} learns a multimodal speech-text embedding space and uses it for zero-shot audio classification. Inspired by this, we hypothesize that multimodal data can benefit the audio-to-intent classification task as well using audio encoders \cite{hsu2021hubert,schneider2019wav2vec} and pre-trained language models \cite{BERT, song2020mpnet}. Additionally, in the context of conversational AI bots, the system is equipped with a set of text sentences provided by the bot developer. These sentences serve as representative examples of each intent that the system can predict. We believe such sentences, if utilized properly, have the potential to achieve generalized zero-shot audio-to-intent classification. We summarize our contributions as follows:
\begin{enumerate}
    \item We propose a \textbf{multimodal learning} approach for simultaneous training of audio and text encoders using a contrastive loss \cite{elizalde2023clap} to align both modalities and a cross-entropy loss for intent classification. 
    \item We explore conditional \textbf{teacher-student learning} approach where the student model mimics the multimodal audio-text representations to improve audio-based intent classifier. 
    \item We introduce \textbf{generalized zero-shot audio-to-intent classification} leveraging a fine-tuned audio encoder from a pre-trained self-supervised model and neural audio synthesizer to create audio embeddings.
\end{enumerate}


\section{Supervised Audio-to-intent classification}
\label{sec:a2i}

In this section, we propose to leverage multimodal audio-to-intent classification model to inject lexical information into an audio-based intent classifier. The multimodal classification problem can be more formally introduced as follows: let $A = \{a_1, a_2, .., a_N\}$ be the set of $N$ audio files, $T = \{t_1, t_2, .., t_N\} $ be the set of text transcripts (paired), and $Y = \{y_1, y_2, .., y_N\}$ be the corresponding output intent labels, such that $y_i \in S$. $S$ defines the set of unique intents seen at training time. We can then define a dataset $D = \{a_i, t_i, y_i\}_{i=1}^N$, where $N$ denotes number of examples in a batch. We use a dataset $D$ to train a model $f$, such that $f(a_i, t_i) = y_i$ for some $i$.



We describe our methods and proposed approaches in the following subsections, where we leverage pre-trained, self-supervised models to extract audio and text representations, which we use to classify speaker's intent using an end-to-end approach.

\vspace{-0.3cm}
\subsection{Multimodal learning}
\label{sec:mm}

\begin{figure}[ht!]
	\begin{minipage}[b]{1.0\linewidth}
		\centering
		\centerline{\includegraphics[width=8.5cm]{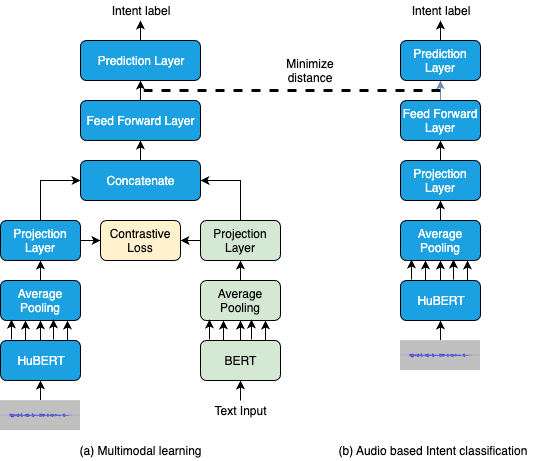}}
	\end{minipage}
	\caption{Architecture of supervised audio-to-intent classification: (a) Multimodal learning using audio and text modalities. (b) Intent classification with audio only modality.}
	\label{fig:a2i_tea_stu}
	\vspace{-0.5cm}
\end{figure}

Figure \ref{fig:a2i_tea_stu}(a) shows the network architecture of our multimodal model. We use HuBERT \cite{hsu2021hubert} as an audio encoder, with which we extract frame-level features from all the utterances in $A$. We then average the features across each utterance over time and represent them as $\widebar{H}_{a}$. We then project $\widebar{H}_{a}$ onto a lower dimension using linear layer, which we denote as $E_{a}$:
\begin{equation}\label{eq:e_a}
    E_{a} = W^{a} * \widebar{H}_{a}  + b^{a},
\end{equation}
where $W^{a}$ and $b^a$ denote the weight matrices and bias of the linear layer, respectively. 

We also use BERT \cite{BERT} as a text encoder, and we convert all the text transcripts in $T$ into a sequence of hidden representations. We average the representations across each transcript over time and represent the result as $\widebar{H}_{t}$. We then project $\widebar{H}_{t}$ onto a lower dimension using a linear layer, which we denote as $E_{t}$:  
\begin{equation}\label{eq:e_t}
    E_{t} = W^{t} * \widebar{H_{t}} + b^{t},
\end{equation}
where $W^{t}$ and $b^t$ denote the weight matrices and bias of the linear layer, respectively. 


\subsubsection{Intent classification}
As shown in figure \ref{fig:a2i_tea_stu}(a), the joint embeddings $E_{at}$ are obtained by concatenating audio representations and text representations followed by a feed forward layer with \textit{ReLu} activation function. These joint embeddings are then passed through a softmax layer to predict the final output $\hat{Y}$:
\begin{align}
    E_{at} &= \text{ReLu}(W^{at}(E_{a} \oplus E_{t}) + b^{at}),\\
    \hat{Y} &= \text{Softmax}(W^y * E_{at} + b^y),
\end{align}
where $W^{at}, W^{y}, b^{at}, b^{y}$ denote weight matrices and bias of the concatenation and classification layer, respectively. The model is optimized using cross entropy loss:
\begin{equation}
    \mathcal{L}_{IC} = -\frac{1}{N} \sum_{j=1}^N \sum_{i=1}^S y_{ij} log(\hat{y}_{ij}),
\end{equation}
where $N$ is the number of samples in a batch, $S$ is the number of target classes, $y_{ij}$ is the ground truth label for the $j^{th}$ sample and $i^{th}$ class, and $\hat{y}_{ij}$ is the predicted probability for the $j^{th}$ sample and $i^{th}$ class.

\subsubsection{Learning a multimodal space}
Drawing inspiration from  \cite{elizalde2023clap,deshmukh2022audio}, we adopt a joint training approach where we simultaneously train audio and text encoders to acquire a shared multimodal space using contrastive loss. From Eq.~\ref{eq:e_a} and Eq.~\ref{eq:e_t}, the audio ($E_{a} \in \mathbb{R} ^ {N \times d}$) and text ($E_{t} \in \mathbb{R} ^ {N \times d}$) representations are in the same dimension. We can compare their similarity as:
\begin{equation*}
    C = \tau * (E_{a} \cdot E_{t}^T),
\end{equation*}
where $\tau$ is a temperature parameter and the similarity matrix $C$ has $N$ correct pairs in the diagonal and $N^2-N$ incorrect pairs in the off-diagonal. 

The model undergoes training utilizing the contrastive loss \cite{elizalde2023clap} paradigm between the audio and text embeddings in pair, following the same loss function in \cite{radford2021learning}:
\begin{equation}\label{eq:l_sim}
    \mathcal{L}_{CL} = 0.5 * (l_{audio}(C) + l_{text}(C)),
\end{equation}
where $l_{k} = CrossEntropy(C)$ along the text and audio axis, respectively. This leads to the total loss:
\begin{equation}
    \mathcal{L}_{MM} = \frac{1}{2} (\mathcal{L}_{IC} + \mathcal{L}_{CL})
\end{equation}
The process of multimodal learning involves training by back-propagating the loss exclusively through the HuBERT model, while keeping the weights of the BERT model fixed. This progressive approach aims to converge the audio embeddings towards the space defined by the text encoder, ultimately leading to the creation of a shared representation space.

\vspace{-0.5cm}
\subsection{Teacher-student learning}
\label{sec:teacher-student}
\vspace{-0.1cm}
In Section \ref{sec:mm}, we anticipate receiving pairs of audio and text inputs. However, during inference, we lack access to both modalities. Our goal is to predict intent when provided with an audio input. As depicted in Figure \ref{fig:a2i_tea_stu}(b), we exclusively train a model with audio data only. To incorporate lexical information from the joint model, we employ conditional teacher-student learning. This approach leverages transfer learning, where a student model learns to emulate the output distribution of a teacher's model.

The student model is trained using audio data only. The HuBERT model takes in audio signals as input and extracts features at the frame level. We then compute the mean of the frame level features and pass it through a linear layer followed by a feed forward layer with \textit{ReLu} activation function to obtain utterance level audio representation $E_{s}$. We then pass $E_{s}$ through a prediction layer that predicts the intent of the audio. The cross entropy loss computed over predicted intent logits and ground truth $(L_{Intent})$ is used as a loss function for backpropagation. 

By employing transfer learning, we train an audio based intent classification model to generate audio embeddings that closely resemble the joint embedding obtained from the multimodal model. This is achieved by minimizing $\ell_2$ norm squared between the two embeddings:
\begin{equation}
    \mathcal{L}_{Student} =  \| E_{at} - E_{s} \|_2^2.
\end{equation}
The overall training objective of the teacher-student model is then defined as the following weighted combination of the cross entropy loss and the $\ell_2$ norm squared:
\begin{equation}
    \mathcal{L}_{Total} = \mathcal{L}_{Intent} + \gamma * \mathcal{L}_{Student}.
\end{equation}

\section{Generalized zero-shot audio-to-intent classification} \label{general-0shot}

\begin{figure}[ht!]
	
	\begin{minipage}[b]{1.0\linewidth}
		\centering
		\centerline{\includegraphics[width=8.5cm, height=9cm]{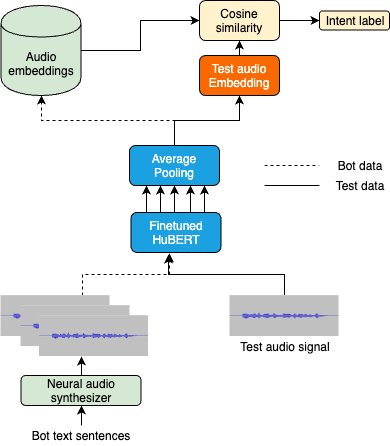}}
	\end{minipage}
	\caption{Generalized zero-shot intent classification utilizing bot text sentences and neural audio synthesizer.}
	\label{fig:gen-zero-shot}
	\vspace{-0.5cm}
\end{figure}

\label{sec:gen-zero}
Within a conversational AI bot system, the bot developer provides sample text sentences along with their corresponding intent labels. However, the corresponding audio samples are not provided. It is possible that the intents presented by the bot developer may not align with the intent set used during our training. Consequently, our objective is to predict the intent of an audio utterance that could be either present in the training set or remain unseen during the training process.

\vspace{-0.4cm}
\subsection{Pre-computation of audio embeddings}
\label{sec:bot_emb}
\vspace{-0.1cm}
To review the generalized zero-shot audio-to-intent classification, we denote $U = \{u_1, u_2, .., u_M\}$ be the set of $M$ text sentences, $V = \{v_1, v_2, .., v_M\}$ be the corresponding intent labels, such that $v_i \in W$, provided by the bot developer. $W$ defines the set of unique intents present in the bot definition. 
Audio embeddings are created by generating an audio file for each text sentence in $U$ using neural audio synthesizer \cite{amazon_polly} as shown in Figure \ref{fig:gen-zero-shot}. Next, we pass each audio file through the HuBERT model and extract frame-level features, followed by average pooling. We associate each audio embedding with its corresponding intent label. This step is conducted only once as an offline process, and all the embeddings are stored in the ``audio embeddings" database. 

\vspace{-0.4cm}
\subsection{Inference}
\label{sec:zero_inf}
\vspace{-0.2cm}
Given a test audio file $Z$, we pass it through the HuBERT model, followed by average pooling. We then compute the cosine similarity between the test audio embeddings and each audio embedding from the embedding database. This similarity calculation helps us determine how closely each audio embedding in the embedding database aligns with the test audio. We identify the utterance in the embedding database that has the highest similarity score to the test audio embedding. We then use the intent label associated with this utterance as the final intent.

\vspace{-0.5cm}
\section{Experiments}
\vspace{-0.3cm}
\label{sec:experiments}

\subsection{Datasets}
\vspace{-0.2cm}
Our proposed intent classification approach is trained and evaluated using various datasets, including publicly available SLURP \cite{bastianelli2020slurp} dataset, as well as audio classification datasets such as AudioMNIST \cite{becker2018interpreting}, TIMIT \cite{garofolo1993timit}, and Speech Commands \cite{warden2018speech}. 

To evaluate zero-shot intent classification, we partition the SLURP dataset, which we refer to as SLURP-Full, into two subsets: SLURP-Seen, which includes 57 intents for the training phase, and SLURP-Unseen, which comprises the remaining unseen 12 intents for the zero-shot evaluation. Additionally, we create another subset, SLURP-mix, consisting of a mix of seen and unseen intents for the generalized zero-shot intent classification task. The SLURP-mix subset consists of 18 intents, of which 12 intents are from SLURP-Unseen and 6 intents are from SLURP-Seen. 

Furthermore, to assess the robustness and generalization capabilities of our modeling strategy, we train the teacher and student models on an internal goal-oriented dialog dataset. This dataset consists of 88 hours of audio data and consists of 90 intents from following domains: “Airline”, “FastFood”, “Finance”, “Health-care”, “Insurance”, “MediaInternetTelecom”, “Travel”, “Retail”, “General”. Statistics about train, development and test splits for SLURP and internal intent datasets are provided in Table \ref{tab:dataset}.


\vspace{-0.5cm}
\begin{table}[th]
    \centering
        \caption{Datasets used for supervised and zero-shot intent classification}
        \vspace{0.7mm}
        \begin{tabular}{ ccccc }
         \toprule
         Name & No.Of. & Train & Dev. & Test \\
         & Intents & & & \\
         \midrule
         Slurp-Full & 69 & 119880 & 8690 & 13078\\
         
         SLURP-Seen & 57 & 111835 & 7859 & 12025\\
         SLURP-Unseen & 12 & 8045 & 832 & 1053 \\
         SLURP-Mix & 18 & 10854 & 1120 & 1589 \\
         Internal & 90 & 57484 & 7219 & 7244 \\
         \bottomrule
        \end{tabular}
    \label{tab:dataset}
    \vspace{-0.5cm}
\end{table}

\vspace{-0.2cm}
\subsection{Models}
\vspace{-0.1cm}
\subsubsection{Baseline models}
\vspace{-0.2cm}
In our research, we benchmark supervised intent classification task using BERT-CTC model \cite{higuchi2022bert}, which predicts intent along with ASR by prepending an intent label to the corresponding output sequence. We also evaluate with partially fine-tuned HuBERT model \cite{hubert_wav2vec}, PF-hbt-large, where GRU based decoder is added to decode intent directly from the fine-tuned HuBERT embeddings. For generalized zero-shot intent classification, we consider HuBERT-Frozen as our baseline model. For zero-shot audio classification tasks, we evaluate our model against the semantically aligned speech-text embedding model as the baseline \cite{huzaifah2023analysis}. 
\vspace{-0.4cm}
\subsubsection{Multimodal teacher model}
\vspace{-0.2cm}
The teacher model represents the audio modality with a HuBERT (hubert\_large\_ll60k) \cite{hsu2021hubert} model, the average-pooled audio embeddings of which are projected using a feed forward layer with an output size of 128. Similarly, the text modality is represented using a BERT-base model, the average-pooled embeddings of which are then projected to a 128-dimensional space using a feed forward layer. The multimodal embeddings, obtained by concatenating the audio and text embeddings, are further transformed using a feed forward layer into a 128-dimensional space. Finally, these embeddings are projected to the known dimension of intent space using a softmax layer. We use a dropout of 0.2 and 0.3 for training the uni-modal projection layers and multimodal feed forward layer, respectively. Our models are trained using Adam optimizer with an initial learning rate set to 0.0001 and a ReduceLROnPlateau learning rate scheduler. The $\tau$ for computing the similarity of audio and text embeddings is set to 0.007. We fine-tune only the top 6 layers of the HuBERT transformer block. This is because top layers are responsible for capturing language and domain-specific information, while the lower layers capture acoustic information \cite{kumar2022investigation}. The BERT model, on the other hand, is frozen. 
\vspace{-0.4cm}
\subsubsection{Student model}
\vspace{-0.2cm}
\label{subsec:stu_model}
Similar to the teacher model, the student model represents the audio modality with a HuBERT model, the average-pooled audio embeddings of which are projected using a feed forward layer with an output size of 128. These projected audio embeddings are further transformed using a feed forward layer and \textit{ReLu} activation function with an output size of 128, which is projected into the known dimension of intent space using a softmax layer. The student model doesn't model any text modality and is trained to minimize the distance between its audio-based embeddings with teacher model's multimodal embeddings as described in Section \ref{sec:teacher-student}. The $\gamma$ for minimizing the distance between joint embedding representation and audio-only embedding representation is set to 10. Student model also fine-tunes top 6 layers of the HuBERT transformer block.

\vspace{-0.3cm}
\section{Results and discussion}
\vspace{-0.2cm}
\subsection{Supervised Intent Classification}
Although the primary focus of this study is zero-shot intent classification, we begin by evaluating the performance of a multimodal (MM) teacher model and its student counterparts in a supervised intent classification task on the SLURP-Full and SLURP-Seen datasets. We have trained a total of five models and each model is described below.
\vspace{-0.2cm}
\begin{itemize}
    \item \textbf{Audio-only}: Model trained using audio-only modality without any guidance from the teacher model.
    \vspace{-0.2cm}
    \item \textbf{MM}: Model trained using audio and text modalities. 
    \vspace{-0.2cm}
    \item \textbf{MM-CL}: Model trained using audio and text modalities with additional contrastive loss.
    \vspace{-0.2cm}
    \item \textbf{Stu-MM}: Model trained using audio-only modality with guidance from the MM teacher.
    \vspace{-0.2cm}
    \item \textbf{Stu-MM-CL}: Model trained using audio-only modality with guidance from the MM-CL teacher.
    
\end{itemize}

The results of these evaluations are presented in Table \ref{tab:sup_ic}. MM and MM-CL models can not be used during inference as we don't have access to the transcript of the audio file. Performance of these models will guide as an upper bound for most of the student and audio-only models. It can be seen that student models (Stu-MM, Stu-MM-CL) outperform baseline BERT-CTC (with K=1) model and audio-only model. The results suggest that student models that are infused with lexical information and trained with contrastive loss are better able to understand the meaning of words, phrases, and the context of an utterance. This improved understanding could help them to perform better on intent classification. To compare our results to those in \cite{hubert_wav2vec}, we also fine-tune all layers of the HuBERT transformer block on the SLURP-Full dataset, as noted in the ``All" column. It can be seen that all of our models outperform PF-hbt-large \cite{hubert_wav2vec} model.

\begin{table}[tp]
    \centering
        \caption{Supervised Intent Classification accuracy on SLURP-Full and SLURP-Seen datasets. \textbf{Top-6}: Top 6 layers of HuBERT transformer block are fine-tuned. \textbf{All}: All layers of HuBERT transformer block are fine-tuned.}
        \vspace{0.7mm}
        \begin{tabular}{ cccc}
         \toprule
         \multirow{3}{*}{Model} & \multicolumn{3}{c}{SLURP}\\
         & \multicolumn{2}{c}{Full} & Seen \\
         \cmidrule{2-4}
         & Top-6 & All & \\
       \midrule
        BERT-CTC \cite{higuchi2022bert} & 87.0 & - & - \\
        PF-hbt-large \cite{hubert_wav2vec} & - & 89.22 & - \\
        
        Audio-only & 85.43 & 89.73 & 84.79 \\
        MM & 90.20 & 90.33 & 91.22 \\
        MM-CL & 90.77 & 90.14 & 91.89 \\
        Stu-MM & 86.26 & \textbf{90.43} & 86.88 \\
        Stu-MM-CL & \textbf{87.21} & 89.49 & \textbf{87.46} \\
         \bottomrule
        \end{tabular}
    \label{tab:sup_ic}
    \vspace{-0.3cm}
\end{table}

\vspace{-0.3cm}
\begin{table}[t]
    \centering
        \caption{Zero-shot Intent classification accuracy on SLURP-Unseen and SLURP-Mix datasets using models trained on SLURP-Seen and Internal intent datasets}
        \vspace{0.7mm}
        \begin{tabular}{ cccc}
         \toprule
         \textbf{Model} & \multirow{2}{*}{\textbf{Model Name}} & \multicolumn{2}{c}{\textbf{SLURP}} \\ 
         \textbf{trained on} &   & Unseen & Mix \\
        \midrule
         & HuB-Frozen & 20.79 & 24.29 \\
        \cmidrule{1-4}
         \multirow{3}{*}{SLURP-Seen} & HuB-FT-AO & 85.38 & 87.48 \\
         & HuB-FT-MM & 87.85 & \textbf{88.86} \\ 
         & HuB-FT-MM-CL & \textbf{88.13} & 88.36 \\
        \cmidrule{1-4}
         \multirow{3}{*}{Internal} & HuB-FT-AO & 28.11 & 28.13 \\
         & HuB-FT-MM & 44.25 & 42.86 \\
         & HuB-FT-MM-CL & \textbf{46.34} & \textbf{45.5} \\
         \bottomrule
        \end{tabular}
    \label{tab:zero}
    \vspace{-0.4cm}
\end{table}

\subsection{Generalized Zero-Shot Intent Classification}

We evaluate zero-shot intent classification by generating audio embeddings from SLURP-Unseen and SLURP-Mix training sentences, as described in Section \ref{sec:bot_emb}. We then evaluate on the SLURP-Unseen and SLURP-Mix test datasets as explained in Section \ref{sec:zero_inf}. We compare four variants of HuBERT models for this task, which are described below. The results are presented in Table \ref{tab:zero}.

\begin{itemize}
    \vspace{-0.2cm}
    \item \textbf{HuB-Frozen}: Pre-trained HuBERT model from HuggingFace \cite{huggingface}
    \vspace{-0.3cm}
    \item \textbf{HuB-FT-AO}: HuBERT model fine-tuned in ``Audio-only" model
    \vspace{-0.3cm}
    \item \textbf{HuB-FT-MM}: HuBERT model fine-tuned in Stu-MM model
    \vspace{-0.3cm}
    \item \textbf{HuB-FT-MM-CL}: HuBERT model fine-tuned in Stu-MM-CL model
    \vspace{-0.2cm}
\end{itemize}

Our analysis indicates that the fine-tuned HuBERT models exhibit a significant performance improvement over the frozen HuBERT model. Furthermore, the student models, HuB-FT-MM and HuB-FT-MM-CL trained using knowledge distillation from the corresponding teacher models outperform the HuB-FT-AO model. This clearly indicates that the lexical infusion and contrastive loss in the teacher model leads to performance improvement. 

Table ~\ref{tab:zero} shows that student models trained using teacher-student learning on the mismatched Internal dataset significantly outperform the HuB-Frozen and HuB-FT-AO models. This demonstrates that our training strategy still achieves superior generalization even when trained on mismatched datasets. However, the performance gap between the models trained on SLURP-Seen and Internal datasets emphasizes the importance of training and fine-tuning the models on a representative dataset encompassing different audio and semantic characteristics to achieve generalization.

\begin{table}[t]
    \centering
        \caption{Evaluation of zero-shot intent classification accuracy on AudioMNIST(AM), Speech Commands (SC) and TIMIT}
        \vspace{0.7mm}
        \begin{tabular}{cccc}
         \toprule
         \textbf{Model} & \textbf{AM} & \textbf{SC} & \textbf{TIMIT} \\
        \midrule
         Baseline \cite{huzaifah2023analysis} & 15.88 & 14.27 & 67.14 \\
         HuB-Frozen & \textbf{32.06} & 25.56 & 52.86 \\
         HuB-FT-AO & 16.69 & 45.87 & 100 \\
         HuB-FT-MM & 26.92 & \textbf{51.26} & 100 \\
         HuB-FT-MM-CL & 29.16 & 46.85 & \textbf{100} \\
         \bottomrule
        \end{tabular}
    \label{tab:zero-baseline}
    \vspace{-0.4cm}
\end{table}

\vspace{-0.3cm}
\subsection{Zero-Shot Audio Classification}
Given that the audio-based models can be applied to zero-shot audio classification using the generalized zero-shot framework introduced in Section \ref{sec:gen-zero}, we proceed to assess and compare these models on TIMIT, AudioMNIST and Speech Commands datasets. We created audio embeddings by synthesizing 10 random SX sentences from TIMIT's Test set, digits 0-9 for AudioMNIST, 10 auxiliary words (“Bed”, “Bird”, “Cat”, “Dog”, “Happy”, “House”, “Marvin”, “Sheila”, “Tree”, and “Wow”) for Speech Commands. Test sets are used as described in \cite{huzaifah2023analysis}. We then evaluated respective test datasets as discussed in Section \ref{sec:zero_inf}. The results of this evaluation are shown in Table \ref{tab:zero-baseline}. Notably, HuB-FT-MM and HuB-FT-MM-CL models demonstrate significant improvement over the baseline models and HuB-FT-AO. The HuB-Frozen model outperforms all other models on the AudioMNIST dataset, potentially due to its exposure to a substantial number of spoken digits during training. We would like to emphasize that our models achieved accuracy of 100\% on the TIMIT dataset. This performance can be attributed to the fact that the linguistic content of the test audio embedding and corresponding audio embeddings in the database are identical.

\vspace{-0.3cm}
\section{Ablation study}
\label{sec:ablation_study}
\vspace{-0.2cm}
\textbf{Validating output from different layers}: We conduct an analysis to understand the behaviour when output from the projection layer or feed forward layer is used to calculate the similarity between test audio embedding and each audio embeddings in the embedding database. In Table \ref{tab:ablation}, we present a comparison of the results obtained from these different layers. Our findings reveal an interesting pattern: as the data progresses through the higher layers of the model, it tends to learn task-specific features, leading to a significant drop in accuracy. 

\vspace{-0.4cm}
\begin{table}[th]
    \centering
        \caption{Zero-shot intent classification using embeddings from Average Pooling layer, projection layer and feed forward (FF) layer.}
        \begin{tabular}{cccc}
         \toprule
         \multirow{2}{*}{\textbf{Model}} & \textbf{Average} & \textbf{Projection } & \multirow{2}{*}{\textbf{FF layer}} \\
          & \textbf{Pooling} & \textbf{Layer} & \\
         \midrule
         HuB-FT-AO & 85.37 & 70.56 & 65.432 \\
         HuB-FT-MM & 87.85 & 76.73 & 70.84 \\
         HuB-FT-MM-CL & 88.12 & 75.499 & 69.136 \\
         \bottomrule
        \end{tabular}
    \label{tab:ablation}
    \vspace{-0.2cm}
\end{table}

\noindent \textbf{Training audio-only model with synthesized data}: We develop an audio-only intent classification model by training it on neural audio synthesizer generated audio files using the SLURP-Unseen training and development datasets. The model's architecture remains consistent with the one described in section \ref{subsec:stu_model}. This model achieves an accuracy of 88.41\% when evaluated on the SLURP-Unseen test data. This result is very close to SLURP-Seen HuB-FT-MM-CL accuracy on SLURP-Unseen test data in the zero-shot intent classification results in Table \ref{tab:zero}.

\vspace{0.2cm}
\noindent \textbf{Zero-shot performance on different sample sizes}: Table \ref{tab:zero} demonstrates significantly improved performance when the number of text sentences are in thousands (8045 for SLURP-Unseen). However, it is important to note that in practical scenarios, bot developers may not have access to such an extensive data. Consequently, we conduct an analysis to investigate the performance of our models using few text sentences from each intent, aiming to provide insights into their behavior under more realistic conditions.

\begin{figure}[htb]
	
	\begin{minipage}[b]{1.0\linewidth}
		\centering
		\centerline{\includegraphics[width=8.5cm]{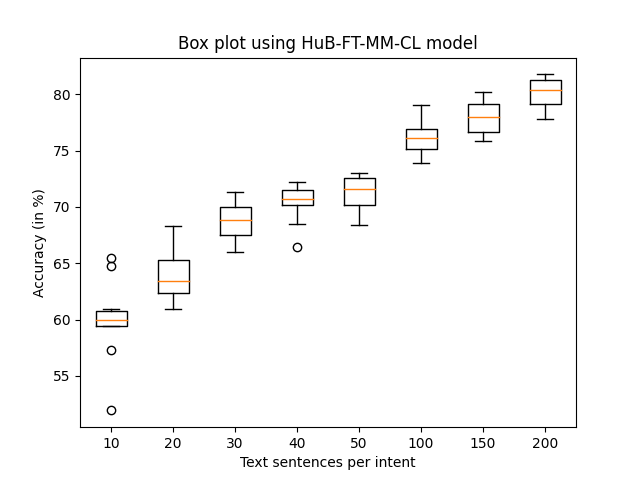}}
	\end{minipage}
	\caption{Accuracy of zero-shot intent classification as size of the text sentences per intent changes}
	\label{fig:sample_sizes}
	\vspace{-0.2cm}
\end{figure}

We choose 10 to 200 random sentences for each intent from the SLURP-Unseen training set and converted them into audio embeddings as described in Section \ref{sec:bot_emb} and evaluated on SLURP-Unseen test set with HuB-FT-MM-CL model. We repeated this experiment 10 times to understand how the model's performance varies with different sets of sentences. Figure \ref{fig:sample_sizes} shows the model's accuracy at different numbers of text sentences per intent. The variation in accuracy is high when the number of sentences per intent is between 10 to 40. Although all models still achieve reasonable accuracy with 10 sentences per intent, the ideal number of sentences per intent is 30, as shown in the graph.

\vspace{-0.2cm}


	

\vspace{-0.6cm}
\section{Conclusion}
\label{sec:page}
\vspace{-0.3cm}
In this work, we propose training a multimodal model using audio and text modalities and contrastive loss. This model is further used to guide a student model by leveraging transfer learning, which captures the distribution of the joint embedding representation. Our results show that student models achieve higher accuracy than state-of-the-art and audio-only audio-to-intent classifiers. We have reported that generalized zero-shot audio-to-intent classification achieves 2.75\% and 18.2\% on unseen intents of SLURP using SLURP seen and internal goal-oriented dialog datasets, respectively. For future work, we aim to investigate the clustering of audio embeddings in zero-shot classification and reducing computation overhead during inference.
\bibliographystyle{IEEEbib}
\bibliography{strings,refs}

\end{document}